# Beyond "Usability and User Experience", Towards an Integrative Heuristic Inspection: from Accessibility to Persuasiveness in the UX Evaluation

## A Case Study on an Insurance Prospecting Tablet Application

Josefina Isabel Gil Urrutia[1, 2,*], Eric Brangier[1],
Véronique Senderowicz[2], and Laurent Cessat[2]

[1]University of Lorraine – Psychologie Ergonomique et Sociale pour l'Expérience Utilisateurs
UFR Sciences Humaines et Sociales, - BP 30309 - Ile du Saulcy - 57006 Metz, France
[2]Allianz Informatique – Direction des Développements et de la Maintenance, Département Digital & Distribution - Tour Franklin 100 Terrasse Boieldieu - BP 1050 La Défense – 92042 Paris La Défense, France

Corresponding author (*): josefina.gil_urrutia@allianz.fr

**Abstract.** Heuristic inspections are often carried out in a rather restrictive manner in the sense that they often address one or two of User Experience aspects. These two generally being: usability and "user experience". This fails to consider UX as it should be [considered]: through a holistic approach. Thus, we suggest to go beyond that by opting for what we have called an Integrative Heuristic Inspection that takes into account issues of: accessibility, usability, emotions & motivation and persuasion, and that aims to simplify the overflow of recommendations UX professionals are faced with nowadays. We illustrate our proposal by a case study carried out on an insurance prospecting tablet application. We analyzed the results of the inspection separately for each dimension as well as combined across dimensions. Implications for a reflection on the structuring of the criteria for a general criteria-based approach in UX are discussed.

## 1 Introduction

A vast number of heuristics has been identified and is regularly used to design and to evaluate human-computer interfaces/interactions (HCI). However, said criteria often lack cohesion. In certain cases, they overlap with one another and/or are interdependent. Thus, they appear to be unnecessarily redundant making heuristic inspections more complex than need be.

After a brief review of the emergence of the four currently recognized main criteria sets [in HCI/UX] and the contributing factors having led to it, this paper aims to argue in favor of a cohesive, integrative revised taxonomy of the existing criteria sets and a derived corresponding model: "the experience-based criteria". We illustrate our argumentation with a case study based off the results from the heuristic inspection of an insurance prospecting tablet application.

## 2 Theoretical Background

Throughout the past 60-70 years, technological developments as well as evolving customer needs/requirements and the transformation of economic models have determined the evolution of the HCI models and dictated the focus of researchers in HCI/UX. This has allowed for new theories, concepts, methodologies, frameworks and different interventional practices to be proposed and adopted in the study of Ergonomics/UX at given times. Indeed, after starting at issues concerning technological accessibility, moving on to ease-of-use and switching to affective factors especially during the rise of video games in the 80's, the past decade has seen the spotlight be put on technological persuasiveness (see Brangier & Bastien [2] for a detailed historical review).

The evolution of the heuristic inspection framework has thus seen four main sets of criteria/guidelines be suggested at given times [2, 4]. These guidelines represent a mean to measure the quality of interfaces and are a valuable reference tool during their design and evaluation. At first, the Accessibility Guidelines were created in order to make information and technology equally accessible – though not strictly – for people with special needs ("design-for-all" approach). As an example, we can name the Web Content Accessibility Guidelines (WCAG) developed by the Web Accessibility Initiative (WAI) of the World Wide Web Consortium (W3C) [16]. Later on, numerous authors addressed issues pertaining to ease-of-use by suggesting Usability Guidelines – such as Ravden & Johnson [12], Nielsen [10] and Bastien & Scapin [1] among others. Usability guidelines aim to adapt the system to the user's and the task's characteristics in order to reduce the probability of errors and increase efficacy, efficiency and user satisfaction. Next, the affective dimensions influencing the interaction [between the system and the user] were addressed. Again, many authors carried out an array of research in this area and proposed their sets of guidelines covering matters such as self-expression, user motivation and aesthetics of the interface [5, 8, 11]. Emotions have been recognized to play an important role in HCI and human intelligence in general [5]. As de Vicente & Pain [5] point out: *"affective states influence [students'] learning and [their] cognitive state in general"*. Lastly, Persuasiveness Guidelines were most recently suggested [9, 12]. These include means to capture the user's attention in order to prolong the interaction with the ultimate goal of changing their attitude(s) and/or behavior(s). They can be applied to various fields such as commercial websites seeking to engage and retain clients; social media platforms aiming to incite people to take part in social movements; and/or e-learning tools to engage and motivate students [3].

This evolution of the HCI framework further evidences how the user experience is based on two main components: the functional experience and the lived experience. This shift of perspective shows how not only accessibility to the information and the system's usability (functional/utilitarian experience: goal-oriented behavior) determine the success - or failure - of an interactive product, but how emotional and persuasive factors (lived experience: quest for rich stimulating and memorable experiences) are also paramount to the user experience. Indeed, affective, motivational and social aspects favour – and are a requirement for – technological acceptance.

Our main concern is the fact that interface design and heuristic inspections are often carried out with a rather restrictive approach in the sense that they often address one, or two, of HCI dimensions. Most commonly, these are the usability and the emotional factors (the latter popularly being referred to as "User Experience"). We consider UX to be a multidimensional systemic phenomenon, and thus we find it is of utmost importance to develop a framework and a method of heuristic design/inspection that acknowledge all aspects of UX.

# 3 Problem

This study falls in line with Brangier et al.'s [4] suggestion to work towards creating an inspection grid that regroups the 4 main dimensions of UX. Accordingly, we have identified four main reasons to work towards suggesting an integrative approach. These are the following:

1. Overlappings/Redundancies: several criteria are present in two of the grids. For example, the Perceivable criterion (accessibility grid) and the Legibility for optimal Guidance criterion (usability grid) both require that a written text meet certain standards for it to be easily read by the user (i.e.: sufficient contrast between the lettering and the background, size of the font, etc.);
2. Cause-effect link: certain criteria appear to have a causal effect on other criteria (i.e.: general ease-of-use determines the persuasiveness of a web site [7];
3. Interdependencies: certain criteria can influence one another such as in what Hassenzahl [8] calls the "interplay of beauty and usability" where the affective dimensions through the aesthetics of the interface might impact its perceived usability (see also [14]);
4. Functional vs. Lived: during the design phase, there is a risk of favoring either the functional or lived experience aspects of the interaction at the expense of the other. For instance, in work applications, the emotional factors are often considered irrelevant and consequently neglected in the design. However, gamification has proven to be a promising alternative for the design of work environments by aiming to engage and motivate workers emphasizing on emotional and persuasive elements.

By acknowledging these remarks, one recognizes the need for a holistic approach of the user experience, which – according to our vision – calls for the ensemble of the dimensions we have considered relevant in UX: accessibility, usability, emotions & motivation, persuasion. Designing(/evaluating) for the user experience with such a selection of grids is a complex task. Our goal here is to illustrate this through a case study of the application of the integrative heuristic inspection, which will serve to further argue towards the simplification of the selection of criteria and the understanding of its organization in the cognitions of UX professionals.

# 4 Case Study

## 4.1 Description

A Native iOS tablet application was developed for the insurance sales agents as a prospecting tool. It's proposed as an alternative work tool that offers mobility and a more attractive tactile interface. The application comprises a variety of functionalities amongst which the possibility to create a "Client File" which was designed based off the equivalent paper version. This is the section this study was carried out on and is what will be referred to as "the app" henceforth.

It consists of 17 consecutive screens that together make up a broad form. The app helps the sales agents to do as much an extensive profile analysis on the prospect/client (P/C) as possible by allowing them to gather any information on the P/C's possessions (such as car(s), home(s), etc.), their contact information, currently owned insurance contracts, professional and family status, etc. as well as to assess their finances and needs in order to accordingly recommend a customer-tailored solution (10 consecutive screens). If the P/C is a business owner, the file also serves to register all of the relevant information (i.e.: monthly fees and taxes, current financial situation, healthcare/retiring/savings plans for the employees, etc.) (next 4 screens). It is solely a client's needs diagnostics and evaluation tool so it does not allow for any quote & buy. The remaining 3 screens serve to log in, to schedule a follow-up appointment and to take note of any client referrals.

Upon completion and validation by the agent, the app generates a PDF that is sent via email to the agent. All of the information is injected into the Client Relationship Management system and into the agent's client portfolio for future consultation/modifications.

## 4.2 Usage

Since the app's release, 1064 sales agents have acquired a tablet and created a user profile as required. Usage data analysis shows that there have been 14,339 sessions, which equals an average of 13.5 sessions per agent (or just over 1session per agent per month). Similarly, an average of 22.9 daily users account for the average 44.5 daily sessions, meaning that each daily user logs in twice approx. The average session duration is of 6 minutes 35 seconds (max=31 minutes and 51 seconds).

The overall stickiness rate is 6%. This means that out of 100 agents who try out the app, 94 of them are just one-time-users and only 6 will reuse the app. Taking this into account and considering that filling out an entire "Client File" takes over an hour, these numbers raise an alarm as to the misuse of the app.

As a result, one might wonder whether the app hasn't encountered much success because of a lack of pragmatic and/or hedonic quality (or for some other reason). Here, we explore this question.

### 4.3 Material & Procedure

A heuristic inspection was carried out by one inspector. It was done using PowerPoint: for every element, a print-screen was taken, resized to actual size (14.9 x 19.87cm) and inserted into a slide on which the annotations were added. The main common structure of the screens of the app, plus the specific content of 6 screens were analysed in depth by applying all 4 sets of criteria.

For our study, we chose to work with the following grids (detailed in Table 1):

- The WCAG 2.0 Guidelines (WAI-W3C) which is based on 4 principles that have also been adapted for non-web Information and Communications Technologies (WCAG2ICT) [15, 16];
- Bastien & Scapin's [1] Usability Grid which is organized around 8 criteria (18 sub-criteria);
- de Vicente & Pain's [5] Motivational Model elements composed of 4 "permanent traits" and 5 "transient states" for the Emotions & Motivation dimension;
- Persuasiveness: Nemery, Brangier & Kopp's [9] grid which includes 8 static and dynamic criteria divided in 23 sub-criteria.

**Table 1.** Criteria per dimension used for the heuristic here (sub-criteria not presented)

| Accessibility | Usability | Emotions & Motivation | Persuasion |
|---|---|---|---|
| Perceivable | Guidance | Control | Credibility |
| Operable | Workload | Challenge | Privacy |
| Understandable | Explicit Control | Independence | Personalization |
| Robust | Adaptability | Fantasy | Attractiveness |
| | Error Management | Confidence | Solicitation |
| | Consistency | Sensory Interest | Priming |
| | Significance of Codes | Cognitive Interest | Commitment |
| | Compatibility | Effort | Ascendency |
| | | Satisfaction | |

Other than the occurrences of each individual error, a pairings analysis was done. This served to identify the overlappings/redundancies, the interdependencies and the cause-effect links between the criteria and particularly the sub-criteria. We also made an effort to classify whether these pairings errors had an impact on the functional or on the lived experience (or both).

## 4.4 Results

Figure 1 illustrates our integrative heuristic inspection approach by showing the ensemble of observations made on a single screen.

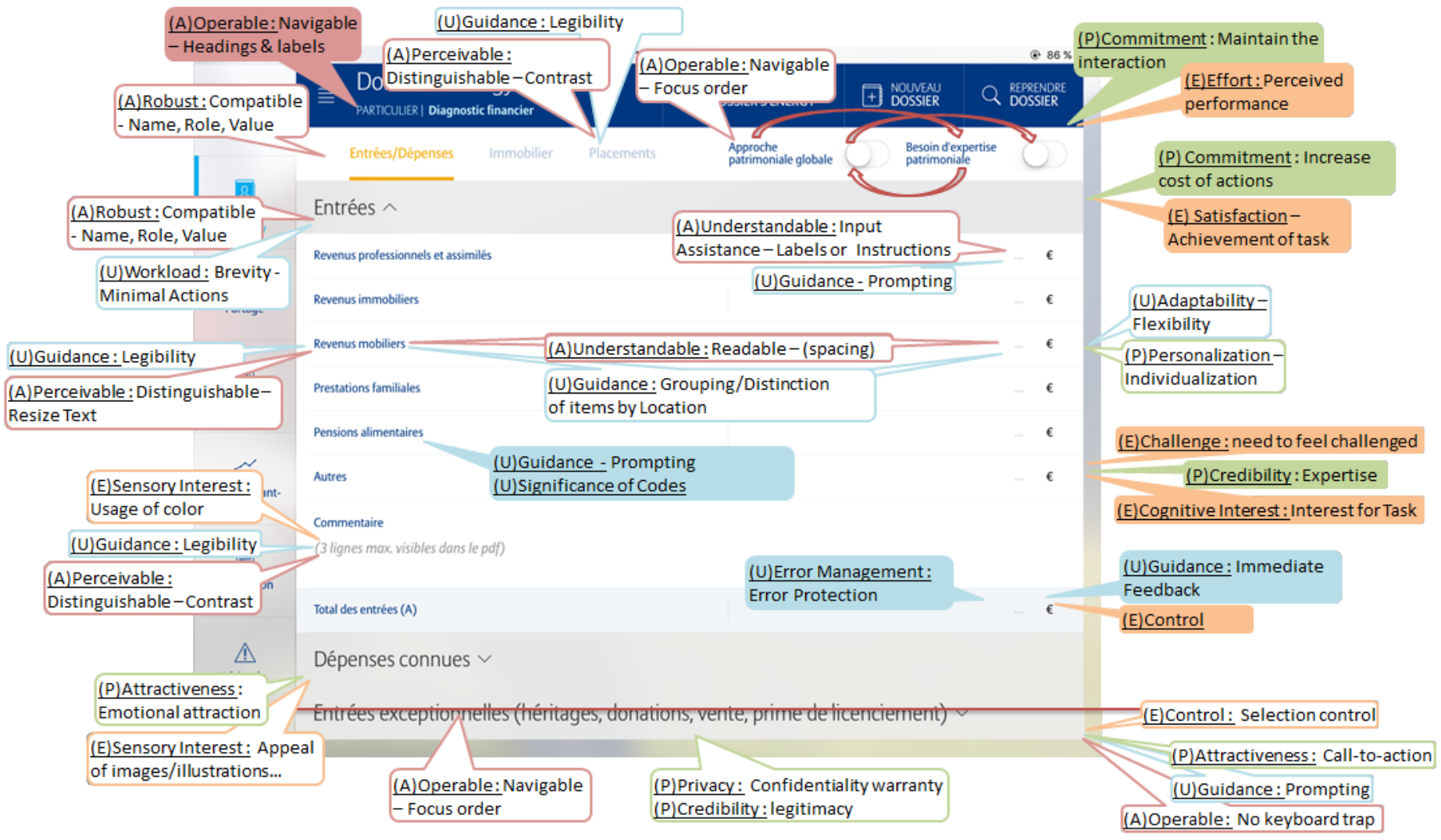

**Fig. 1.** Illustration of the Integrative Heuristic Inspection by applying all 4 sets of criteria to the analysis of one screen[1].

The ensemble of errors detected on all seven inspected elements are reported in the Appendix.

**Analysis by dimension.** At first, we inspected the app dimension by dimension.

*Accessibility*. The app failed to meet WCAG's standards concerning all four criteria. Numerous interface components hinder the app's perceivability, especially the poor legibility of textual elements, as well as its operability because of many keyboard traps and illogical focus navigation order that cause the loss of meaning. Moreover, elements that should help with error identification and a better understanding of labels and instructions are often insufficient (i.e. signaling an error by simply highlighting the field with a pale yellow color) or missing. The app's general compatibility with assistive technologies is also rather weak.

[1] Legend: in red/marked (A) Accessibility criteria; in blue/marked (U) Usability criteria; in orange/marked (E) Emotions & Motivation criteria; in green/marked (P) Persuasiveness criteria. White background = criteria not met vs. Colored background = criteria met.

*Usability.* We found no issue with the explicit control guideline, and only rare, minor errors regarding the consistency and the significance of codes principles. In addition, except for a few design choices that prolong the interaction unnecessarily (mishaps regarding minimal actions), the workload principle is also satisfactorily met: labels are concise and there is no information overload. Concerning error management: though error correction remains possible as long as the file hasn't been closed, protection against error occurrence and especially the quality of the error messages could be improved. Furthermore, the app doesn't offer any means to customize the interface depending on personal preferences or needs, and certain elements that aren't relevant to the task cannot be deactivated (i.e. the main menu remains visible at all times and cannot be hidden). However, the app allows the user to fill out the form in any order they might wish to (flexibility). It also facilitates the navigation through the screens either by sliding through the screens (sequential order), or through the table-of-contents-like lateral collapsible menu.

The two criteria that were least met are guidance and compatibility. Numerous prompting errors were found, the most important one being the absence of a call-to-action that would suggest the user to *slide* to the next screen in order to continue filling out the form (instead of stopping at the last item of the screen). Another common mistake is the format given to the interactive fields: the user needs to *tap* on a specific area (marked: "…") for it to become active. In addition, legibility is often hindered due to text format (font color-background contrast, font size, etc.). Only a few minor errors pertaining to the distinction/grouping of items by format or by location were identified and, because of the app's reactivity (in spite of some occasional delays), the immediate feedback criterion was equally met. Finally, with regard to compatibility, the digital version in comparison to the original paper version isn't optimal. Other than a few non-negligible differences between the two versions relative to changes in the structure of the form and the order of the elements, certain design choices for the digital version prolong the process of filling out the file. This is due to its segmentation in 16 screens (vs 3 recto-verso A4 pages) and the determined means of navigation, which make jumping from one section to another more challenging (instead of just having to relocate and/or adjust the arm/wrist in order to be able to write).

*Emotions & Motivation.* The affective dimension was particularly difficult to evaluate with this method and with the grid we chose. No elements pertaining to the criteria confidence, independence and fantasy were found. Nonetheless, we identified numerous elements that could be a potential source of frustration for the user due to the blockages they cause, possibly resulting in a feeling of loss of control. Only on the most important screens, we found that the general interface might pose a challenging stimulating situation for the user.

Infrequent use of illustrative images or aesthetic elements as well as the use of dull colors results in an interface that would not seem to evoke much sensory interest for the user. Likewise, only the screens corresponding to key elements of the file (most important information that should be obtained from the P/C), might trigger interest for the task and stimulate the user on a cognitive level – but again, these are limited cases. Finally, for the most part, the app lacks an indicator of progression that could serve the user as a mean to measure their achievement and provide them with a sense of satisfaction. Only the most important screens would seem to procure this feeling.

*Persuasiveness.* The app inspires credibility through proof of fidelity (buttons' labels are true to their associated actions) and expertise (the company can be recognized through the look & feel of the app and the logo on the log in screen). However, it fails to do so by demanding certain personal data of the P/C without justifying why. Similarly, a confidentiality warranty notice is only available on the log in screen (privacy).

Additionally, a more pronounced use of aesthetics and attractive elements such as catchy call-to-actions or illustrative images could reinforce the app's attractiveness. Moreover, certain elements of the interface might be seen as external negative factors that ought to be avoided when filling out a file and interfere with user commitment (such as the main menu of the main application that is non-collapsible). On the other hand, other interface elements on the most important screens help to maintain the interaction through a stimulating increasing level of task difficulty.

Lastly, the ascendency criterion did not apply to this context; there were only a few positive elements regarding the priming and solicitation principles, and no personalization elements were identified (customization is not possible, and group membership is only visible on the log in screen).

**Pairings/Cross-Grid Analysis.** Then, by combining the observations made, we were able to identify the overlappings, cause-effect links and interdependencies across the criteria/sub-criteria. Table 2 recaps these occurrences as well as the results of the classification of the impact on functional and/or lived experience for the majority of the misses (criterion non-met (-)) and hits (criterion met (+)) found. Below, we mention certain of these cases.

**Table 2.** Overall occurrences of cross-dimensions hits and misses.

| | Misses (-) | Hits (+) |
|---|---|---|
| Overlappings | 33 | 1 |
| Cause-effect link | 15 | 8 |
| Interdependencies | 26 | 8 |
| Impact: | | |
| Functional: | 47 | 2 |
| Lived: | 7 | 15 |

*Overlappings.* 33 cases of cross-grid criteria/subcriteria overlappings were found. Some of these cases are:

- (-) Use of 2 formats to distinguish between only two options (Distinction by Format, Guidance; Usability) is also non-compliant with the criterion use of color (Distinguishable; Accessibility)

- (-) "…" is non-representative of the information that's expected (Labels/Instructions, Input Assistance, Understandable; Accessibility ; Prompting, Guidance; Usability)

*Cause-effect link.* 15 cause-effect-like links were found. For instance:

- (+) Usage of illustrative images stimulates sensory interest (Emotions & Motivation), and thus encourages emotional attraction (attractiveness; Persuasion) favoring a positive lived experience
- (-) Lack of a distinctive mark signaling that an information is mandatory results in the occurrence of an error and causes the user to feel that they have lost control over the system (Error Identification, Input Assistance, Understandable; Accessibility ; Error Protection, Error Management; Usability ; Order of Actions, Control; Emotions&Motivation)
- (-) Using terms that are too vague (significance of codes; Usability) induces a lack of fidelity-credibility for the interface (Persuasion)

*Interdependencies.* 26 cases of criteria/sub-criteria interdependencies were identified. Some examples are:

- (+) Usage of properly prompting text favors Usability and is indicative of the fidelity (Credibility, Persuasion) of the action associated with a certain call-to-action
- (-) Displaying non-relevant elements for the task at hand as required by the error protection principle (error management; Usability) doesn't allow for the avoidance of external negative factors thus hindering user commitment (Persuasion)

## 5 Discussion & Conclusion

This case study shows how an interface can be analyzed through at least four lenses with regard to four different approaches that have occupied the study of HCI/UX throughout time. These four approaches correspond to four dimensions of user experience that we consider relevant and important to date. We carried out what we call an integrative heuristic inspection in order to argue in favor of the merging of these four dimensions by suggesting the combination of the different criteria sets allowing for a holistic approach to HCI and most importantly to UX.

Other than singular errors, through this study we identified numerous cases of redundancies among the criteria/sub-criteria across the dimensions. Also, we identified cases where not meeting a certain criterion caused that another criterion from another grid was also not satisfied (cause-link effect). Lastly, we found cases of interdependencies between 2 criteria, or even more.

From our point of view, this supports previous works that encourage UX professionals to adopt a holistic approach in the field. Indeed, user experience is the result of the interaction of a set of factors that together explain the success (acceptance) or failure (rejection) of a technology. Likewise, surpassed by the ever-growing amount of recommendations (i.e. criteria), this study also backs our objective of conducting a reflection as to the structuring of said criteria set aiming partly to propose a simplification of its classification.

Having said that, these four dimensions fall short to consider all of the possible aspects that contribute to UX. We find it fails to take into account elements pertaining to culture, economic factors as well as socio-organizational factors. Future research should investigate this.

Also, we know of only one previous study that carried out an inspection similar to the one presented here. The authors [6] applied a similar method for the evaluation of a proactive informational driving system. Broadening the scope of application by means of more case studies within a wider selection of interactive products and across different domains (i.e. social media, games) would represent valuable data and elucidate a better understanding of this issue.

**Acknowledgments.** We thank Stéphane Lermechain and Kien Nguyen for their contributions to this study.

# Appendix

Overall errors detected through the inspection across all 7 inspected elements. Empty columns and rows have been removed.

| | | | | Persuasion | | | | | Emotions & Motivation | | | | | Usability | | | | | | | | | | |
|---|---|---|---|---|---|---|---|---|---|---|---|---|---|---|---|---|---|---|---|---|---|---|---|---|
| | | | | Static criteria | | | Dynamic criteria | | Permanent Traits | | Transient States | | | Guidance | | | | Workload | Adaptability* | Error Management | | | | |
| | | | | Credibility | Attractiveness | | Priming | Commitment | Control | | Sensory Interest | Cognitive Interest | Satisfaction | | Grouping/Distinction of Items | | | Brevity | | | | | | |
| | | | | Fidelity. | Emotional attraction | Call-to-action | Encouragements | factor avoidance | Selection control | Order of Actions | sounds, videos… | Compatibility. | Achievement of task. | Prompting(*) | by Location (*) | by Format (*) | Legibility (*) | Minimal Actions (*) | Flexibility (*) | Error Protection (*) | Quality of Error Messages (*) | Consistency (*) | Compatibility (*) | **Individual Errors** |
| Accessibility | Perceivable | Text alternatives | to non-text content | | | | | | | | | | | | | | | | | | | | 1 | |
| | | Adaptable | Information, Structure | | | | | | | | | | | 1 | | | | | | | | | | |
| | | | Meaningful sequence | | | | | | | | | | | 1 | | | | | | | | | | |
| | | | Sensory characteristics | | | | | | | | | | | | | | | | | 1 | | | | 2 |
| | | Distinguishable | Use of color | | | | | | | | | | | | | 1 | | | | | 4 | | | |
| | | | Contrast | | | | | | | | 2 | | | | | | 11 | | | | | | | |
| | | | Resize text | | | | | | | | | | | | | | 7 | | | | | | | |
| | | | Visual Presentation | | | | | | | | | | | | | | | | | | | | | 2 |
| | Operable | Keyboard accessi | Keyboard | | | 1 | | | 1 | | | | | 1 | | | | | | | | | | |
| | | | No keyboard trap | | | 5 | | | 4 | | | | | 5 | | | | | | | | | | 2 |
| | | Navigable | Bypass blocks | | | | | | | | | | | | | | | | 1 | | | | | |
| | | | Focus order | | | 9 | | | 4 | | 5 | | | 11 | | | | 1 | | | | | | 7 |
| | | | Focus Visible | | | | | | | | | | | | | | | | | | | | | 1 |
| | Understandable | Readable | Language of page/parts | | | | | | | | | | | | 3 | | | | | | | | | |
| | | Input Assistance | Labels or instructions | | | | | | 1 | | | | | 9 | | 1 | | | | | | | | |
| | | | Error identification | | | | | | | 3 | | | | | | | | | | 5 | | | | |
| | | | Error suggestion | | | | | | | | | | | | | | | | | 2 | | | | |
| | Robust | Compatible | Name, Role, Value | | | | | 1 | | | | | | | | | | | | 1 | | 2 | | 6 |
| Usability | Guidance | Prompting (*) | | | | 14 | | | 14 | | 5 | | | | | | | | | | | | | 1 |
| | | Grouping/Distinction | by Location (*) | | | | | | | | | | | | | | | | | | | | | 2 |
| | | | by Format (*) | | | | | | 1 | | | | | | | | | | | | | | | |
| | | Immediate Feedback (*) | | | | | 1 | | | | | | 1 | | | | | | | | | | | 1 |
| | | Legibility (*) | | | | | | | | | 2 | | | | | | | | | | | | | |
| | Workload | Brevity | Minimal Actions (*) | | | | | | 2 | | | | | | | | | | | | | | | 12 |
| | Explicit Contr | Explicit User Actions (*) | | | | | | | | | | | | | | | | | | | | | | |
| | | User Control (*) | | | | | | | | | | | | | | | | | | | | | | 1 |
| | Adaptability | Flexibility (*) | | | | | | | | | | | | | | | | | | | | | | |
| | | User Experience | | | | | | | | | | | | | | | | | | | | | | |
| | Error Managemen | Error Protection (*) | | | | | | 1 | | 3 | | | | | | | | | | | | | | 2 |
| | | Quality of Error Messages (*) | | | | | | | | | | | | | | | | | | | | | | 1 |
| | Consistency (*) | | | | | | | | | | | | | | | | | | | | | | | 4 |
| | Signifiance of Codes (*) | | | 4 | | | | | | | | | | | | | | | | | | | | |
| | Compatibility (*) | | | | | | | | | | | 1 | | | | | | | | | | | | 7 |
| Emotions & Motivation | (Permanent | Control | Selection control | | | 6 | | | | | | | | | | | | | | | | | | |
| | (Transient states) | Sensory Interest | Appeal of graphs, sounds, | | 3 | 5 | | | | | | | | | | | | | | | | | | |
| | | Effort | Perceived performance | | | | | | | | | | | | | | | | | | | | | 1 |
| | | Satisfaction | Achievement of task | | | | 1 | | | | | | | | | | | | | | | | | |
| | | | Measure of achievement | | | | | | | | | | | | | | | | | | | | | 1 |
| Persuasion | Static criteria | Credibility | Expertise. | | | | | | | | | | | | | | | | | | | | | 2 |
| | | | Legitimacy. | | | | | | | | | | | | | | | | | | | | | 2 |
| | | Privacy | Safety | | | | | | | | | | | | | | | | | | | | | 1 |
| | | | Confidentiality warranty | | | | | | | | | | | | | | | | | | | | | 6 |
| | | Personnalisation | Individualization. | | | | | | | | | | | | | | | | | | | | | 2 |
| | | | Group membership | | | | | | | | | | | | | | | | | | | | | 1 |
| | Dynamic criteria | Solicitation | Suggestion | | | | | | | | | | | | | | | | | | | | | 1 |
| | | | Teasing | | | | | | | | | | | | | | | | | | | | | 1 |